\documentclass[conference]{IEEEtran}

\IEEEoverridecommandlockouts

\newif\ifshort
\shortfalse 

\usepackage{cite}
\usepackage{amsmath,amssymb,amsfonts}
\usepackage{algorithmic}
\usepackage{graphicx}
\usepackage{textcomp}
\usepackage{xcolor}

\usepackage{enumitem,listings,caption,booktabs}
\usepackage[hidelinks]{hyperref}
\usepackage[normalem]{ulem}
\usepackage[T1]{fontenc}
\usepackage[scaled=0.85]{beramono}

\newcommand*{\RefS}[1]{\hyperref[#1]{\S\ref*{#1}}}
\newcommand*{\RefSection}[1]{\hyperref[#1]{Section~\ref*{#1}}}
\newcommand*{\RefAppendix}[1]{\hyperref[#1]{Appendix \ref*{#1}}}

\newcommand*{\RefRemark}[1]{\hyperref[remark:#1]{Remark #1}}
\newcommand*{\RefInvar}[1]{\hyperref[#1]{Invariant \ref*{#1}}}
\newcommand*{\RefLemma}[1]{\hyperref[lemma:#1]{Lemma #1}}

\newcommand*{\smallcaps}[1]{\small\textsc{#1}\normalsize}
\newcommand*{\subsubtitle}[1]{\vspace{2pt}\textbf{#1:}}

\newcommand*{\RefAlgorithm}[1]{\hyperref[#1]{Algorithm \ref*{#1}}}

\ifshort
\newcommand*{\RefAlgTitle}[1]{\lstinline{#1}}
\else
\newcommand*{\RefAlgTitle}[1]{\hyperref[alg:#1]{\lstinline{#1}}}
\fi

\newcommand{\extendedurl}[0]{\url{https://arxiv.org/abs/1234.12345}}

\lstdefinelanguage{Go2}{
        morekeywords=[1]{
                        break,default,func,interface,select,case,defer,go,map,vid
                        struct,chan,else,goto,package,switch,const,fallthrough,
                        if,range,in,type, continue,for,import,return,var,while, and, or, xor, not},
        morekeywords=[2]{
                        append,cap,close,complex,copy,delete,imag,
                        len,make,new,panic,print,println,real,recover,
                        assert},
        morekeywords=[3]{
                        bool,byte,complex64,complex128,error,float,float32,float64,
                        int,int8,int16,int32,int64,rune,string,
                        uint,uint8,uint16,uint32,uint64,uintptr},
        morekeywords=[4]{true,false,iota,nil},
        morestring=[b]{"},
        morestring=[b]{'},
        morestring=[b]{`},
        comment=[l]{//},
        morecomment=[s]{/*}{*/},
        sensitive=true
}

\DeclareCaptionFormat{lstcaptionformat}{\hrule\small\vspace{1pt}\hspace{2pt}\textit{#1#2#3}}
\DeclareCaptionFormat{lstcaptionformatnoalg}{\hrule\small\vspace{1pt}\hspace{2pt}\textit{#3}}
\begin{document}

\title{
    Maximum Flow on Highly Dynamic Graphs
    \thanks{
        This work was supported by the Natural Sciences and Engineering Research Council of Canada (NSERC).
        
        \ifshort An extended version of this paper is available on arXiv:\\\extendedurl\fi{}
    }
}

\author{
\IEEEauthorblockN{Juntong Luo}
\IEEEauthorblockA{\textit{The University of British Columbia} \\
Vancouver, Canada \\
luuo2000@ece.ubc.ca}
\and
\IEEEauthorblockN{Scott Sallinen}
\IEEEauthorblockA{\textit{The University of British Columbia} \\
Vancouver, Canada \\
scotts@ece.ubc.ca}
\and
\IEEEauthorblockN{Matei Ripeanu}
\IEEEauthorblockA{\textit{The University of British Columbia} \\
Vancouver, Canada \\
matei@ece.ubc.ca}
}

\maketitle
\unless\ifshort
\thispagestyle{plain}
\pagestyle{plain}
\fi

\begin{abstract}
Recent advances in dynamic graph processing have enabled the analysis of highly dynamic graphs with change at rates as high as millions of edge changes per second. Solutions in this domain, however, have been demonstrated only for relatively simple algorithms like PageRank, breadth-first search, and connected components. Expanding beyond this, we explore the maximum flow problem, a fundamental, yet more complex problem, in graph analytics. We propose a novel, distributed algorithm for max-flow on dynamic graphs, and implement it on top of an asynchronous vertex-centric abstraction. We show that our algorithm can process both additions and deletions of vertices and edges efficiently at scale on fast-evolving graphs, and provide a comprehensive analysis by evaluating, in addition to throughput, two criteria that are important when applied to real-world problems: result latency and solution stability.
\end{abstract}


\section{Introduction}

\noindent
\textit{Graph streaming} systems aim to ingest an evolving graph as a stream of graph updates possibly arriving at a high rate, and repeatedly provide, on-demand or regularly, results to a standing graph analytics query. Such systems have received increasing attention in recent years due to the prevalence of applications involving social networks, communication networks, financial transactions, and other dynamic systems \cite{besta2020practice, aggarwal2014evolutionary, mcgregor2014graph}.

To obtain good performance on massive graphs, these graph streaming frameworks are \textit{parallel} \cite{graphone, graphbolt}, and sometimes \textit{distributed} \cite{sallinen2016graph, sallinen2019incremental}. However, in this domain, only a few problems have \textit{algorithmic} solutions that can harness a parallel or a distributed platform: the few existing algorithms solve relatively simple problems like PageRank, breadth-first search, and connected components \cite{besta2020practice, mcgregor2014graph}. Algorithms for other problems, especially more complex ones, remain largely unexplored. 

This paper proposes a novel algorithm for a more complex problem -- maximum flow. Given a graph with capacities on each edge, a source vertex $s$, and a sink vertex $t$, the maximum flow problem asks for the maximum amount of flow allowed from $s$ to $t$. This problem and its dual, minimum cut, are fundamental problems in the field of network flow and graph theory \cite{west2001introduction}. They have a wide range of applications \cite{Ahuja1993network}, including transportation \cite{harris1955fundamentals}, communication \cite{homa2015efficient}, web community identification \cite{flake2000efficient, imafuji2004finding}, link spam detection \cite{saito2007large}, image segmentation \cite{jensen2022review, boykov2001experimental}, online voting systems \cite{tran2009sybil}, and others \cite{krishnamurthy1984improved, hong2004distributed}, many of which operate on real-world data that is \textit{dynamic} in nature.

While there exist a few parallel and distributed algorithms to track maximum flow on dynamic graphs \cite{zhou1996self, ghosh1997self, hong2004distributed, pham2006adaptive, homa2011asynchronous, homa2015efficient, khatri2022scaling, luo2023going}, none of them are able to achieve all of the following three goals: \textit{(i)} handle both vertex and edge updates, \textit{(ii)} handle both additions and deletions efficiently, and \textit{(iii)} handle graphs that evolve quickly (\RefSection{sec:related-work:max-flow}). This paper proposes a parallel/distributed algorithm that meets these three goals.

Our algorithm targets a vertex-centric, shared-nothing, and asynchronous dynamic graph processing model free of constraints like batching topology updates (\RefSection{sec:model}). 
This model has been shown to be scalable, enables \textit{real-time} analysis by allowing graph analytic query results to be extracted on-demand with arbitrary granularity and low latency, and offers more than an order of magnitude higher performance for frequent queries compared to prior work \cite{sallinen2016graph, sallinen2019incremental, sallinen2023realtime}. 

This paper offers the following contributions:
\vspace{-3pt}
\begin{itemize}[leftmargin=*]

    \item We propose a dynamic maximum flow algorithm%
        \unless\ifshort\footnote{
        Our earlier previous workshop paper~\cite{luo2023going}, proposed a preliminary version of this algorithm, proved its correctness, and demonstrated orders of magnitude speed-ups over a \textit{static}, snapshot-based, approach for frequent queries. However, this preliminary version has low performance when facing frequent deletions and lacks critical performance optimizations. Here we use a new mechanism to handle deletions much more efficiently, add critical optimizations, and provide a thorough evaluation at scale. The previous paper evaluated the algorithm on random sources and sinks, whereas here we evaluate the new algorithm on hard cases, i.e., using the the most popular vertices as sources and sinks.
    }\fi%
    ~targeting highly dynamic real-world graphs (\RefSection{sec:algorithm}). Our algorithm efficiently supports additions and deletions of vertices and edges at high rates, as well as on-demand queries. Our algorithm matches well parallel and distributed platforms, and is robust to asynchronicity and concurrent graph updates.
    \ifshort\vspace{4pt}\fi
    
    \item We provide a comprehensive evaluation on real-world graphs with up to hundreds of millions of edges of evolution (\RefSection{sec:evaluation}). Specifically, we test extreme cases with the most popular vertices being sources and sinks, and evaluate the following aspects: \textit{(i)} \textit{throughput} -- maintaining the dynamic solution for a max-flow query with incoming event rates up to millions of changes per second; \textit{(ii)} \textit{scalability} -- being able to compute max flow on dynamic graphs as large as a half-billion edges on a single commodity machine, and further show excellent strong scaling with increasing parallelism; \textit{(iii)} \textit{performance on delete-heavy workloads} -- high-variability in evolution behaviour has minimal impact on our algorithm to provide a quick solution; \textit{(iv)} \textit{result latency} for low ingestion rates -- sub-second query response times; and \textit{(v)} \textit{solution stability} -- unlike static solutions, only minor adjustments are required to move to the latest max flow path as the graph evolves.

\end{itemize}

\section{\label{sec:background}Maximum Flow Background}

\subsection{The Maximum Flow Problem}
\noindent
Given a directed graph $G=(V,E)$ with a capacity $c(u,v)$ for each edge $(u,v)\in E$, the maximum flow problem asks for a flow from the source vertex $s\in V$ to the sink vertex $t\in V$ with the maximum possible value of flow \cite{ford1956maximal}. We define the terms flow and maximum flow below. For convenience, we refer to vertices except $s$ or $t$ as \textit{normal} vertices.

\subsubtitle{Flow}
A \textit{flow} $f \colon E \rightarrow \mathbb{R}$ meets the following conditions:
\vspace{-1pt}
\begin{itemize}[leftmargin=*]
    \item \textit{Capacity constraint}: $f(u,v)\leq c(u,v)$. The flow that passes through an edge cannot exceed the capacity of that edge.
    \item \textit{Flow conservation}: For a normal vertex, the sum of incoming flow must equal the sum of outgoing flow, i.e., it only forwards flow, and does not create or destroy flow.
\end{itemize}

\subsubtitle{Maximum Flow}
A \textit{maximum flow} is a flow with the maximum possible value, which can be computed as the total amount of flow leaving $s$, or equivalently, the amount entering $t$. Note that there may be multiple flows in the same graph that achieve the maximum possible value.

\subsection{Algorithmic Insights}
\noindent
There are two concepts that commonly appear in the context of maximum flow:

\subsubtitle{Residual Graph}
Given a graph $G$ and a flow $f$, the \textit{residual graph} $G_f$ indicates how much additional flow could be sent across edges. For an edge $(u,v)$, the \textit{residual capacity} $c_f(u,v)$ is defined to be the amount of additional flow that can be sent across $(u,v)$, i.e., $c_f(u,v)=c(u,v)-f(u,v)$. An edge $(u,v)$ is \textit{saturated} if $c_f(u,v)\leq0$.

When $f(u,v)$ increases by $\Delta$, the residual graph $G_f$ changes accordingly: \textit{(i)} $c_f(u,v)$ decreases by $\Delta$, and \textit{(ii)} $c_f(v,u)$ increases by $\Delta$. $c_f(v,u)$ increases because flows are \textit{skew-symmetric}, i.e., $f(u,v)=-f(v,u)$. Intuitively, we can view this as allowing $v$ to send back up to $\Delta$ additional units of flow to $u$ by ``returning'' some flow sent from $u$ to $v$ earlier.

\subsubtitle{Augmenting Path}
Given a graph $G$ and a flow $f$, an \textit{augmenting path} is a path from $s$ to $t$ in the induced residual graph $G_f$ where each edge $(u,v)$ has $c_f(u,v)>0$. In other words, it is a path which can be followed to send more flow from $s$ to $t$. Note that $f$ is a maximum flow if and only if there is no augmenting path in $G_f$.

\subsection{Solutions on Static Graphs}\label{sec:background:static-solutions}
\noindent 
Most maximum flow algorithms can be classified into two categories: \textit{(i)} Ford-Fulkerson \cite{ford1956maximal} or \textit{(ii)} push-relabel (also referred to as preflow-push or Goldberg-Tarjan) \cite{goldberg1988new}.

The \textit{Ford-Fulkerson} method proceeds by repeatedly finding augmenting paths in $G_f$ and augmenting flow along the paths until no augmenting paths exist \cite{ford1956maximal}. It maintains a valid flow that eventually evolves into a maximum flow. Notably, this method operates on global graph state, which makes it unsuitable for distributed implementations.

\label{p:pr-bg}The \textit{push-relabel} method, however, maintains a \textit{preflow} and gradually converts the preflow into a valid flow, under the guidance of each vertex's \textit{height} \cite{goldberg1988new}. It ensures no augmenting path exists, and thus the final flow is maximal. We provide a brief overview of this algorithm.

\subsubtitle{Preflow and Excess}
A \textit{preflow} is similar to a flow, but with the flow conservation constraint relaxed. For a vertex $v$, the \textit{excess}, $e(v)$, is the sum of incoming flow minus the sum of outgoing flow. The flow conservation constraint is satisfied and the preflow is a valid flow if and only if the excess is 0 for all normal vertices. 

\subsubtitle{Height}
Each vertex is assigned a \textit{height}, denoted by $h(v)$. When $h(v)<|V|$, then $h(v)$ is an estimate of the shortest path distance to $t$ in $G_f$. When $h(v)>|V|$, then $h(v)-|V|$ is an estimate of the distance to $s$. While the heights of $s$ and $t$ are fixed, i.e., $h(s)=|V|$ and $h(t)=0$, the heights of other vertices start from 0 and may increase.

The push-relabel algorithm proceeds by letting each vertex \textit{push} flow to its neighbours, under the principle of ``flow can only go downhill''.
At the start of the algorithm, the source vertex $s$ generates excess flow and saturates its outgoing edges, making its neighbours \textit{active} (i.e., have non-zero excess flow). Then, \textit{active} vertices repeatedly perform the \textit{push} operation to transfer some excess flow to a neighbour in $G_f$ with a strictly lower height. 
When an active vertex has no valid neighbour to discharge excess flow, it executes the \textit{lift} (also referred to as \textit{relabel}) operation to increase its height to the minimum possible value such that a valid target exists. When a flow cannot reach the sink $t$, it eventually gets pushed back to $s$ when vertices gain sufficient heights.
The algorithm terminates when no active vertex except $s$ and $t$ exists.

The \textit{generic push-relabel algorithm} does not specify how to select an active vertex for the push and lift operations and has a time complexity of $O(V^2E)$. A variant using a \textit{dynamic tree} data structure achieved $O(VE\log(V^2/E))$ \cite{goldberg1988new}. Additionally, it has been shown that using the \textit{global relabeling} heuristic is essential for good practical performance \cite{cherkassky1997implementing}.

We design our parallel/distributed solution for dynamic graphs based on the push-relabel approach, as it is vertex-centric and local: it repeatedly performs the \textit{push} and \textit{lift} operations on each active vertex, and these operations only require local vertex properties and the heights of neighbours.

\section{Computational Model\label{sec:model}}
\noindent Our solution is built on a vertex-centric, asynchronous, and shared-nothing abstraction that preserves scalability and enables solutions to be extracted on-demand with arbitrary granularity and low latency. This section presents an overview of this abstraction, and the reader is referred to \cite{sallinen2023realtime} for a more detailed description.

\subsection{Graph Representation}
\noindent 
We model a graph as a set of vertices. Each vertex maintains a list of its outgoing edges, and can be addressed via a unique identifier. Key to this graph representation is that no state is shared between vertices, so that a graph can be partitioned over a set of compute machines.

\subsection{Processing Model}
\noindent 
We use an asynchronous vertex-centric processing model similar to \smallcaps{HavoqGT} \cite{reza2020highly, pearce2013scaling}: algorithms are designed from the perspective of a vertex that reacts to topology change events or to algorithmic messages received from other vertices. Each vertex is an independent agent. Vertices communicate only through asynchronous (one-way) messages delivered in FIFO order, and there is no shared state or explicit synchronization between them. Multiple vertices may act simultaneously.

Topological change events and algorithmic messages are processed concurrently, with topological events being prioritized. Upon receiving a topology event or a message, a vertex triggers an associated user-defined handler to update its present state and possibly send new messages to other vertices. These messages further propagate algorithmic information and trigger handlers at their destinations. The algorithm proceeds in this recursive manner and terminates when no unprocessed message or event exists.

\subsection{Event Ingestion}
\noindent 
The system begins with an empty graph and ingests a stream of graph topology \textit{events} (e.g., edge add/delete/modify). For each event, the system updates the dynamic graph store and presents the event to the algorithmic layer. As the system aims to handle real-world event streams, it has no a priori information about the future (i.e., no assumption is made about the future graph evolution). For vertex deletion, we expect a vertex to have all of its edges removed before itself being removed.

\subsection{Programming Interface}
\noindent 
The algorithm is expressed as a set of user-defined handlers for messages or topology change events received at a vertex. The interface enables the user to define \textit{(i)} custom data types associated with vertices, edges, and messages, and \textit{(ii)}, as described below, how a vertex reacts to messages and events:
\begin{itemize}[leftmargin=*]
    \item \textit{onMessageReceived:} defines the actions of a vertex when it receives an algorithmic message from another vertex.
    \item \textit{onVertexChanged:} defines the actions of a vertex when itself (excluding its edges) is changed. It may initialize itself, its properties, and/or send messages to other vertices.
    \item \textit{onEdgeChanged:} defines the actions of a vertex \textit{after} an outgoing edge is added, deleted, or updated.
\end{itemize}

\section{\label{sec:algorithm}Push-Relabel for Dynamic Graphs}
\noindent
Our algorithm is based on the static push-relabel algorithm \cite{goldberg1988new} and is designed with the goals of \textit{(i)} handling both additions and deletions of vertices and edges efficiently, and \textit{(ii)} scaling well on parallel and distributed platforms.
Our implementation of the algorithm is open-source\footnote{\url{https://github.com/ScottSallinen/lollipop}}.

\ifshort
This section presents the vertex-local invariants that guide the design of our algorithm (\RefSection{sec:alg-invariant}) and an overview of the algorithm (\RefSection{sec:alg-overview}). Please refer to the extended version of this paper\footnote{\extendedurl} for algorithm details and optimizations.
\else
This section is structured as follows. First, we present the vertex-local invariants that guide the design of our algorithm (\RefSection{sec:alg-invariant}). Then we provide an overview of the algorithm (\RefSection{sec:alg-overview}). Following that, we describe the vertex properties (\RefSection{sec:alg-vertex-prop}) and message format (\RefSection{sec:alg-msg-format}). Finally, we discuss each operation in detail (\RefSection{sec:alg-details}) and the optimizations that are important for good practical performance (\RefSection{sec:alg-optimizations}).
\fi

\subsection{Vertex-Local Invariants\label{sec:alg-invariant}}
\noindent
To design a dynamic algorithm on top of our vertex-centric, asynchronous, and shared-nothing computational model, we follow a \textit{vertex-local invariant restoration} approach, whereby each vertex aims to restore a set of invariants on its \textit{local} properties. This approach provides the advantage that the resulting algorithm can be efficiently implemented on our shared-nothing model (as all invariants are local) and is intrinsically robust to asynchronicity and concurrent graph updates (as each vertex operates independently).

The invariants, as listed below, are derived from the correctness proof of the original push-relabel algorithm \cite{goldberg1988new}.

\begin{enumerate}[leftmargin=*,label*=\textit{\arabic*.}, ref=\arabic*]
    \item \textit{Flow constraints}
          \begin{enumerate}[leftmargin=*,label*=\textit{\alph*.}, ref=1\alph*]
              \item \label{invar:rescap} The residual capacity of an edge is non-negative.
              \item \label{invar:zero-excess} The excess of a normal vertex is $0$.
          \end{enumerate}
    \item \textit{Maximality of flow}
          \begin{enumerate}[leftmargin=*,label*=\textit{\alph*.}, ref=2\alph*]
              \item \label{invar:st-height} The height of the sink is $0$, and the height of the source is no smaller than $|V|$.
              \item \label{invar:rescap-height} For any two neighbouring vertices $v$ and $w$, if $c_f(v,w)>0$, then $h(v) \leq h(w) + 1$. Note that, to achieve this, vertex $v$ needs to track $h(w)$ while there is no shared state between vertices.
          \end{enumerate}
\end{enumerate}

We show these \textit{local} invariants lead to a maximum flow -- a \textit{global} property:

\subsubtitle{Flow} The capacity and flow conservation constraints are satisfied by \RefInvar{invar:rescap} and \RefInvar{invar:zero-excess}, respectively. Since both constraints are met, the final output is a valid flow.

\subsubtitle{Maximum Flow} We show there exists no augmenting path in $G_f$ by contradiction: assume, to the contrary, that such an augmenting path exists, i.e., there exists a path with length $l$ from $s$ to $t$ in $G_f$ with every edge on the path $(v,w)$ having $c_f(v,w)>0$. By \RefInvar{invar:rescap-height}, it follows that $h(s) \leq h(t) + l$. However, since $h(s)\geq|V|$ and $h(t)=0$ (by \RefInvar{invar:st-height}), we have $|V| \leq h(s) \leq l$, which is a contradiction with the maximum possible length of a path in the graph. Therefore, the final output is a maximum flow.

\subsection{Algorithm Overview\label{sec:alg-overview}}
\noindent For static graphs, our algorithm computes a maximum flow in a way similar to the generic push-relabel algorithm (\RefSection{p:pr-bg}, \cite{goldberg1988new}). Therefore, we focus on the scenarios that are unique to dynamic graphs here. Note we handle edge changes indirectly as capacity increases or decreases.

\subsubsection{Vertex Addition} Instead of the individual vertex add events, our algorithm reacts to the event of \textit{an increase in the maximum number of vertices the graph ever has} (\RefAlgTitle{OnNewMaxVertexCount}). For this event, our algorithm increases $h(s)$ to ensure \RefInvar{invar:st-height} holds, 
and let $s$ \RefAlgTitle{push} flow to neighbours if possible. This ensures flow does not return to $s$ before it explores all potential paths to $t$.

\subsubsection{Vertex Deletion} Our computational model expects a vertex to have all of its edges removed before itself being removed. No action is required.

\subsubsection{Capacity Increase} For an increase in $c(v,w)$, we need to ensure it does not create a new augmenting path (i.e., we should maintain \RefInvar{invar:rescap-height}). The algorithm achieves this by lowering $h(v)$ if needed with \RefAlgTitle{restoreHeightInvariant}. Recursively, $v$'s upstream vertices will also decrease their heights due to \RefInvar{invar:rescap-height}. Then, as ``flow goes downhill'', flow from other vertices and $s$ will be pushed towards $v$.%
\footnote{Some other dynamic maximum flow algorithms take a different approach: they force $v$ to send $w$ flow to saturate $(v,w)$, leaving a deficit (i.e., negative excess) at $v$ \cite{goldberg2015faster, khatri2022scaling}.}

\subsubsection{Capacity Decrease} If a decrease in capacity results in a negative residual capacity $c_f(v,w)<0$ on an edge $(v,w)$, the algorithm must ensure \RefInvar{invar:rescap}, $c_f(v,w)\geq0$, is restored. The algorithm achieves this by forcing $w$ to send $-c_f(v,w)$ units of flow to $v$. This might leave a deficit at $w$ ($e(w)<0$), but will increase $c_f(v,w)$ to $0$ and restore \RefInvar{invar:rescap}.


\subsubsection{\label{sec:alg-overview-negative}Negative Excess and Flow} When a vertex $v$ has a negative excess $e(v)<0$, the original push-relabel algorithm is unable to resolve this negative excess. Our previous solution handles this by simply setting $h(v)=-|V|$, resulting in low performance \cite{luo2023going}. To handle negative excess efficiently, for each vertex $v$, our new algorithm manages a second height, named \textit{negative height}, denoted by $h_-(v)$. The algorithm ensures $h_-(t)\geq|V|$ and $h_-(s)=0$, and lets vertices \RefAlgTitle{push} negative flow to $s$ and $t$ under the guidance of $h_-(v)$. Similar to positive flow, negative flow can only be pushed ``downhill'', and $h_-$ are subject to an invariant: for any $v\in V \setminus \{t\} $ and $w\in V$, if $c_f(w, v)>0$, then $h_-(v) \leq h_-(w)+1$.

Apart from negative excess, another problematic scenario on fully dynamic graphs (i.e., graphs with both adds and deletes) is a subgraph that has both positive flow and negative flow but is disconnected from both $s$ and $t$. Due to flow conservation, the amount of positive excess and negative excess is equal and should be cancelled out. However, this scenario creates two issues: \textit{(i)} the positive excess and negative excess could chase each other indefinitely (e.g., in a cycle), and, \textit{(ii)} the active vertices in the subgraph are untouched during global relabeling\unless\ifshort~(\RefSection{sec:alg-optimizations-global-relabeling})\fi, creating either a performance issue or a correctness issue depending on the design of the global relabeling process.

To tackle this problem, our algorithm sets $h(v)=0$ if $e(v)<0$. This resolves the issues as follows: \textit{(i)} when global relabeling is triggered, all vertices in the subgraph get $h_-(v)=\infty$ (as they are disconnected from $s$ and $t$), so the negative excess cannot be pushed away; \textit{(ii)} because $h(v)=0$ when $e(v)<0$, vertices with negative excess will pull positive flow from other vertices. As negative excess stays unmoved and pulls positive flow from others, the positive excess and the negative excess will eventually meet and cancel out.\unless\ifshort\footnote{Note two key differences compared to the pull-relabel algorithm proposed by Khatri et al. \cite{khatri2022scaling}: \textit{(i)} our algorithm pushes negative flow \textit{concurrently} with the positive flow, while their algorithm has to push these two types of flow \textit{in turn} due to sharing the same heights; \textit{(ii)} their global relabeling process does not consider subgraphs that have active flow but are disconnected from both $s$ and $t$, potentially resulting in longer convergence time.}\fi

\unless\ifshort

\subsection{\label{sec:alg-vertex-prop}Vertex Properties}
\vspace{-2pt}
\noindent
\autoref{alg:vertex-prop} shows the properties stored along each vertex. A vertex's \lstinline{Type} can be either \lstinline{Source}, \lstinline{Sink}, or \lstinline{Normal}. \lstinline{Excess} is the difference between the sum of incoming flow and the sum of outgoing flow. \lstinline{HeightPos} ($h$) and \lstinline{HeightNeg} ($h_-$) are the heights for positive flow and negative flow, respectively. The dictionary \lstinline{ResCapOut} stores the residual capacities of the edges from the vertex to its neighbours (we do not track flow on individual edges explicitly).

As the solution should use no shared state between vertices, each vertex also records the \lstinline{HeightPos}, \lstinline{HeightNeg}, and \lstinline{ResCapOut} of its in-neighbours and out-neighbours. Note that tracking neighbour properties is not required under shared-memory assumptions, and could be avoided to gain further performance improvements within such platforms.

For a new vertex, all properties are $0$ or empty by default. We set the source's \lstinline{HeightPos} to $|V|$ and the sink's \lstinline{HeightNeg} to $|V|$ when the vertex is added.

\begin{lstlisting}[caption=Vertex Properties, label={alg:vertex-prop}]
    // Vertex's own properties
    Type      VertexType     // Source, Sink, Normal
    Excess    int            // Incoming flow - outgoing flow
    HeightPos int            // Height for positive flow
    HeightNeg int            // Height for negative flow
    ResCapOut map<uint, int> // Residual capacities to nbrs

    // Tracking neighbour properties
    NbrHeightPos map<uint, int> // Neighbours' HeightPos
    NbrHeightNeg map<uint, int> // Neighbours' HeightNeg
    ResCapIn map<uint, int> // Residual capacities from nbrs
\end{lstlisting}

\subsection{\label{sec:alg-msg-format}Message Format} 
\vspace{-2pt}
\noindent
As shown in \autoref{alg:msg-format}, each message contains the vertex sender ID, the amount of flow, the height for positive flow, and the height for negative flow. A vertex $v$ calls \lstinline{v.send(w, f)} to send a message with \lstinline{f} amount of \lstinline{Flow} to $w$. The \lstinline{SenderId}, \lstinline{HeightPos}, and \lstinline{HeightNeg} are automatically populated.

\begin{lstlisting}[caption=Message Format, label={alg:msg-format}]
    SenderId  uint // Sender ID
    Flow      int  // Amount of flow
    HeightPos int  // Sender's HeightPos
    HeightNeg int  // Sender's HeightNeg
\end{lstlisting}
\vspace{-2pt}

\subsection{\label{sec:alg-details}Algorithm Details}
\vspace{-2pt}
\noindent
We first describe the core operations in the algorithm (1-5), and then describe how a vertex reacts to topological events (6, 7) and other vertices' messages (8).

\subsubsection{push}
The \RefAlgTitle{push} operation attempts to push as much flow (positive or negative) as possible to the specified vertex.
\begin{lstlisting}[caption=push, label={alg:push}]
func push(v *Vertex, w uint) {
    amount := 0
    if v.Excess > 0 and v.HeightPos > v.NbrHeightPos[w]:
        amount = min(v.Excess, max(v.ResCapOut[w], 0))
    else if v.Excess < 0 and v.HeightNeg > v.NbrHeightNeg[w]:
        amount = -min(-v.Excess, max(v.ResCapIn[w], 0))
    if amount != 0:
        v.Excess -= amount
        v.ResCapOut[w] -= amount
        v.ResCapIn[w] += amount
        v.send(w, amount)
}
\end{lstlisting}

\subsubsection{lift}
The \RefAlgTitle{lift} operation lifts the vertex to the \textit{minimum} possible height such that the vertex has a valid target to push some amount of flow. It can only run on normal vertices as $s$ and $t$ have fixed heights. A vertex always has edges with sufficient residual capacities to unload all excess flow. Note that \RefAlgTitle{lift} does not break \RefInvar{invar:rescap-height}.

The \lstinline{liftPos} operation presented below is for a vertex with positive excess, and the operation for one with negative excess is similar (replace \lstinline{NbrHeightPos}, \lstinline{ResCapOut}, and \lstinline{HeightPos} with \lstinline{NbrHeightNeg}, \lstinline{ResCapInt}, and \lstinline{HeightNeg} respectively).
\begin{lstlisting}[caption=lift, label={alg:lift}]
func liftPos(v *Vertex) {         // For liftNeg:
    assert v.Type == Normal and v.Excess > 0  // v.Excess < 0
    minHeight := $\infty$
    for nbrId, nbrHeight in v.NbrHeightPos: // v.NbrHeightNeg
        if v.ResCapOut[nbrId] > 0:       // v.ResCapIn[nbrId]
            minHeight = min(minHeight, nbrHeight)
    assert minHeight != $\infty$
    v.HeightPos = minHeight+1                  // v.HeightNeg
}
\end{lstlisting}

\subsubsection{discharge} 
The \RefAlgTitle{discharge} operation attempts to drain the vertex's excess. For a normal vertex $v$, it repeatedly runs \RefAlgTitle{push} on all neighbours and \RefAlgTitle{lift}s $v$ until no excess is left. For $s$ and $t$, because of their fixed heights, \RefAlgTitle{discharge}returns after attempting all neighbours once. As mentioned in \RefSection{sec:alg-overview-negative}, vertices with negative excess in a subgraph disconnected from $s$ and $t$ (vertices with $h_-(v)=\infty$) cannot push flow away.
\begin{lstlisting}[caption=discharge, label={alg:discharge}]
func discharge(v *Vertex) {
    if v.Excess < 0 and v.HeightNeg == $\infty$:
        return
    while v.Excess != 0:
        for nbrId, nbrHeight in v.NbrHeightPos:
            push(v, nbrId)
        if v.Type != Normal or v.Excess == 0:
            break
        lift(v) // calls liftPos or liftNeg
}
\end{lstlisting}

\subsubsection{restoreHeightInvariant}
This operation restores \RefInvar{invar:rescap-height} between the vertex $v$ and a neighbour $w$. It first attempts to saturate the edge $(v,w)$ by \RefAlgTitle{push}ing flow to $w$ (note $s$ has sufficient excess to saturate outgoing edges). Then, if the invariant is still violated, it descends $v$ to restore the invariant.
\begin{lstlisting}[caption=restoreHeightInvariant, label={alg:restoreHeightInvariant}]
func restoreHeightInvariant(v *Vertex, w uint) {
    push(v, w)
    if v.Type != Normal:
        return
    maxHeightPos := v.NbrHeightPos[w]+1
    if v.ResCapOut[w] > 0 and v.HeightPos > maxHeightPos:
        v.HeightPos = maxHeightPos
    maxHeightNeg := v.NbrHeightNeg[w]+1
    if v.ResCapIn[w] > 0 and v.HeightNeg > maxHeightNeg:
        v.HeightNeg = maxHeightNeg
}
\end{lstlisting}

\subsubsection{\label{alg:broadcastHeightIfNeeded}broadcastHeightIfNeeded}
To ensure each vertex's view of its neighbours is accurate, a vertex must call this operation after it is updated. This operation sends the vertex's new heights to its neighbours if the heights have been changed. 

\begin{lstlisting}[caption=broadcastHeightIfNeeded]
func broadcastHeightIfNeeded(v *Vertex) {
    if heightChanged(v):
        for nbrId, _ in v.NbrHeightPos:
            // heights are automatically populated
            v.send(nbrId, 0)
}
\end{lstlisting}

\subsubsection{onVertexChanged}
Within this event, and when the maximum number of vertices the graph has ever had increased, we instruct $s$ and $t$ to execute a separate event, \RefAlgTitle{OnNewMaxVertexCount}.
This event first increases the heights of $s$ and $t$ to ensure \RefInvar{invar:st-height} is met. As the height is increased, new opportunities for $s$ and $t$ to push flow might appear. Therefore, $s$ and $t$ run the \RefAlgTitle{discharge} operation. Finally, $s$ and $t$ execute \RefAlgTitle{broadcastHeightIfNeeded} to keep their neighbours up-to-date about their heights.
\begin{lstlisting}[caption=OnNewMaxVertexCount, label={alg:OnNewMaxVertexCount}]
func OnNewMaxVertexCount(v *Vertex, newCount uint) {
    if v.Type == Source:
        v.HeightPos = newCount
        discharge(v)
    else if v.Type == Sink:
        v.HeightNeg = newCount
        discharge(v)
    broadcastHeightIfNeeded(v)
}
\end{lstlisting}

\subsubsection{onEdgeChanged}
After an edge $(v,w)$ is added/removed/changed, the capacity across this edge has changed; as such, vertex $v$ calls the \RefAlgTitle{onEdgeChanged} event handler with the change in capacity (referred to as \lstinline{delta}). The handler ignores self-loops (where $v=w$), edges to $s$, and edges from $t$ because they have no effect on the maximum flow. For other changes, the handler performs the following tasks on $v$:
\begin{itemize}[leftmargin=*]
    \item If $w$ is a new neighbour, send a message to $w$.
    \item Update the residual capacity to account for the change.
    \item If $v$ is the source, update the excess. The purpose is to ensure the source has sufficient excess to saturate all its outgoing edges as it is responsible for creating the preflow.
    \item Notify $w$ about this change in $c(v,w)$.
    \item Ensure \RefInvar{invar:rescap-height} holds between $v$ and $w$.
    \item Discharge and broadcast heights to neighbours if needed.
\end{itemize}
\begin{lstlisting}[caption=onEdgeChanged, label={alg:onEdgeChanged}]
func onEdgeChanged(v *Vertex, w uint, delta int) {
    if v == w or isSource(w) or v.Type == Sink:
        return // ignore loops, edges to s, and edges from t
    if not v.NbrHeightPos.contains(w): // new neighbour
        v.NbrHeightPos[w] = 0
        v.NbrHeightNeg[w] = 0
        v.ResCapOut[w] = 0
        v.ResCapIn[w] = 0
        v.send(w, 0) // notify new neighbour
    v.ResCapOut[w] += delta
    if v.Type == Source:
        v.Excess += delta
    v.sendCapacityOffset(w, delta)
    restoreHeightInvariant(v, w)

    discharge(v)
    broadcastHeightIfNeeded(v)
}
\end{lstlisting}

\subsubsection{onMessageReceived}
This handles the event of a vertex $v$ receiving a message from another vertex $w$. The handler performs the following tasks:
\begin{itemize}[leftmargin=*, listparindent=\parindent]
    \item If $w$ is a new neighbour, inform $w$ of $v$'s height.
    \item Stores $w$'s heights in \lstinline{v.NbrHeightPos} and \lstinline{v.NbrHeightNeg}.
    \item If the message intends to inform a change in $c(w,v)$, update \lstinline{v.ResCapIn[w]} accordingly. Otherwise, handle the \lstinline{Flow} in the message by updating $c_f(v,w)$, $c_f(w,v)$, and $e(v)$.
    \item If $c_f(w,v)$ ends up being less than $0$, send sufficient flow to restore $c_f(w,v)\geq0$ (\RefInvar{invar:rescap}).
    \item Ensure \RefInvar{invar:rescap-height} is restored.
    \item If $v$ ends up having a deficit in excess, set \lstinline{v.HeightPos} to $0$, as described in \RefSection{sec:alg-overview-negative}.
    \item Finally, \RefAlgTitle{discharge} the vertex's excess and broadcast heights to neighbours if needed.
\end{itemize}
\begin{lstlisting}[caption=onMessageReceived, label={alg:onMessageReceived}]
func onMessageReceived(v *Vertex, m Message) {
    if not v.NbrHeightPos.contains(m.SenderId): // New nbr
        v.ResCapOut[m.SenderId] = 0
        v.ResCapIn[m.SenderId] = 0
        v.send(m.SenderId, 0)
    v.NbrHeightPos[m.SenderId] = m.HeightPos
    v.NbrHeightNeg[m.SenderId] = m.HeightNeg

    if m.IsCapacityOffset():
        v.ResCapIn[m.SenderId] += m.CapacityOffset
    else:
        v.ResCapOut[m.SenderId] += m.Flow
        v.ResCapIn[m.SenderId] -= m.Flow
        v.Excess += amount

    if v.ResCapIn[m.SenderId] < 0:
        flow := -v.ResCapIn[m.SenderId]
        v.Excess -= flow
        v.ResCapOut[m.SenderId] -= flow
        v.ResCapIn[m.SenderId] += flow
        v.send(s, flow)
    
    restoreHeightInvariant(v, m.SenderId)
    
    if v.Excess < 0 and v.HeightPos > 0:
        v.HeightPos = 0
    discharge(v)
    broadcastHeightIfNeeded(v)
}
\end{lstlisting}

\subsection{\label{sec:alg-optimizations}Optimizations}
\vspace{-2pt}
\noindent
Our baseline algorithm is able to produce correct maximum flow solutions on dynamic graphs, yet, similar to other push-relabel-based algorithms, has a relatively high worst-case time complexity and requires optimizations for good practical performance on large graphs. In this subsection, we present key optimizations we find that greatly improve the performance; first we introduce those proposed in our prior work (1, 2, and some of 3) \cite{luo2023going}, then describe critical new optimizations (4-7).

\subsubsection{\label{sec:alg-optimizations-initial-height}Initial Height}
When a new vertex is given a height of zero, it may lower the heights of its upstream vertices and pull flow from them, even when it has no path to the sink. We avoid this undesirable situation by setting the initial height to $\infty$ and letting the vertex naturally descend itself when restoring \RefInvar{invar:rescap-height} with a neighbour.

\subsubsection{\label{sec:alg-optimizations-overestimate-vertex-count}Projected Vertex Count}
The \RefAlgTitle{OnNewMaxVertexCount} handler is called every time the historical maximum number of vertices $N_\text{max}$ has increased. This is expensive, as it causes flows to be pushed back and forth near the source. To reduce the frequency of calling this handler, we maintain a \textit{projected number of vertices}, $N_p$. When $N_\text{max} > N_p$, we update $N_p$ to $\alpha N_\text{max}$, where $\alpha>1$, and call the handler with the new $N_p$. We set $\alpha=1.1$ in this paper.

\subsubsection{\label{sec:alg-optimizations-global-relabeling}Global Relabeling}
As described in \RefSection{sec:background:static-solutions}, the height $h(v)$ of a vertex $v$ is an \textit{estimate} of $v$'s shortest path distance to $s$ or $t$ (similar for $h_-(v)$). While $h(v)$ and $h_-(v)$ do not \textit{overestimate} the actual distance due to \RefInvar{invar:rescap-height}, they often \textit{underestimates}, as they are not promptly increased when edges on the shortest path become saturated. This has a major impact on the performance \cite{cherkassky1997implementing}, as it misleads where excess flow should be pushed, and correcting it requires flow being pushed back and forth to lift the vertices. To resolve this issue, static solutions perform \textit{global relabeling} (GR) \cite{goldberg1988new} periodically to adjust all vertices to the optimal heights.

We incorporate the heuristic into our dynamic solution, as follows.

\subsubtitle{Triggering Condition} Most existing static push-relabel algorithms (and our previous version) perform GR periodically after a number of lift operations. Yet, we find this simple condition often results in extended runtime in parallel settings. In particular, towards the end of the computation, due to a relatively small number of active vertices (low parallelism) \cite{kulkarni2009much}, heights are adjusted slowly, and GR becomes more important in accelerating convergence. However, the low parallelism may lead to less frequent GR, as the interval is tied to how many times the heights are adjusted (i.e., the number of lifts). To solve this problem, our algorithm employs an additional triggering condition: it also triggers GR when the \textit{time} since the last GR exceeds a threshold, which is a function of the runtime of the last GR. This ensures GR runs promptly regardless of the available parallelism.

\subsubtitle{Height Adjustments} To perform a GR after it is triggered, we introduce four phases in our algorithm: 
\vspace{-2pt}
\begin{itemize}[leftmargin=*, listparindent=\parindent]
    \item $P_\text{normal}$: This is the initial phase. During this phase, our algorithm proceeds as described previously. After GR is triggered, the algorithm enters $P_\text{drain}$.
    \item $P_\text{drain}$: This phase drains all in-flight messages. Vertices cannot lift their heights. When all messages are processed, the algorithm enters $P_\text{relabel-up}$.
    \item $P_\text{relabel-up}$: In this phase, vertices change their positive and negative heights to $\infty$, except for the following vertices, whose heights are set to different values: \textit{(i)} $s$, with heights $h(s)=|V|$ and $h_-(s)=0$, \textit{(ii)} $t$, with heights $h(t)=0$ and $h_-(t)=|V|$, and \textit{(iii)} $v$ with $e(v)<0$, with heights $h(v)=0$ and $h_-(v)=\infty$. During this phase, all vertex operations are disabled. Upon completion, the algorithm enters $P_\text{relabel-down}$.
    \item $P_\text{relabel-down}$: During this phase, vertices restore \RefInvar{invar:rescap-height}. The algorithm achieves this by letting $s$, $t$, and vertices with $e(v)<0$ broadcast their heights to their neighbours, and take advantage of \RefAlgTitle{restoreHeightInvariant} to let vertices descend until \RefInvar{invar:rescap-height} is met globally. The \RefAlgTitle{push}operation is disabled in this phase. Upon convergence (all messages are processed), the algorithm transitions back to $P_\text{normal}$ and resumes normal execution.
\end{itemize}

\subsubsection{Skip Sending Unneeded Heights}
A vertex $v$ tracks a neighbour $w$'s height $h(w)$ for two purposes: \textit{(i)} to decide if $v$ can send flow to $w$, and \textit{(ii)} to ensure $h(v)\leq h(w)+1$ when $c_f(v,w)>0$ (\RefInvar{invar:rescap-height}). Therefore, when $c_f(v,w)\leq0$, $h(w)$ is unimportant for $v$. To reduce messages sent between vertices, we update the algorithm to stop sending $h(w)$ from $w$ to $v$ if $c_f(v,w)\leq0$, and resume sending it when $c_f(v,w)>0$. Similarly, $v$ skips sending $h_-(v)$ to $w$ if $c_f(v,w)\leq0$, as $w$ cannot send negative flow to $v$ when $c_f(v,w)\leq0$.

\subsubsection{Aggregated Operations}
Another optimization that significantly reduces the number of messages is combining \RefAlgTitle{discharge} and \RefAlgTitle{broadcastHeightIfNeeded} across multiple messages. Specifically, because these two operations are not bound to individual messages/events, we can skip them if we know more messages/events will arrive at the vertex. (This optimization requires support from the underlying framework.)

\subsubsection{Optimizing Discharge and Lift}
For vertices with many neighbours, iterating over their neighbour lists is costly (some vertices have more than 10 million neighbours). Since the \RefAlgTitle{lift} operation is only used in \RefAlgTitle{discharge} and both require iterating over the list of neighbours, we combine these two operations to reduce the runtime of processing individual messages.

\subsubsection{Replacing Hash Tables with Arrays}
We also reduce the time it takes to process individual messages by using arrays instead of hash tables. Specifically, a vertex stores neighbour properties in an array, and each message carries the index of the sender in the receiver's neighbour array. To avoid duplicated neighbours, a vertex also maintains a hash table that maps each neighbour's ID to an index in the array. This optimization avoids an expensive hash table lookup for every message and significantly improves the performance.

\fi

\section{Evaluation\label{sec:evaluation}}
\noindent
We evaluate our algorithm on large real-world dynamic graphs from the following aspects: \textit{(i) performance and scalability} -- that is, whether it sustains a high throughput (\RefSection{sec:evaluation:staturation}), scales well with the number of cores (\RefSection{sec:evaluation:scalability}), and handles deletions efficiently (\RefSection{sec:evaluation:deletion}); \textit{(ii) effective resource utilization} -- that is, whether it is able to improve response latency at lower event rates (\RefSection{sec:evaluation:rate-limiting}); and \textit{(iii) solution stability} -- that is, whether it provides solutions that do not vary much as the graph evolves (\RefSection{sec:evaluation:stability}).

A detailed comparison highlighting the benefits of our dynamic solution over a static (i.e., snapshot-based) solution is included in our previous workshop paper \cite{luo2023going}. The fact that the dynamic approach pays off after just a few queries can also be seen indirectly from the plots presented in \RefSection{sec:evaluation:scalability} and \RefSection{sec:evaluation:stability}. 

\subsubsection{Implementation} We implemented our algorithm on top of \textsc{Lollipop} \cite{sallinen2023realtime}, a framework that supports the computational model described in \RefSection{sec:model}. \textsc{Lollipop} is designed to support fast prototyping of dynamic graph algorithms, as it emulates a distributed setting on a single machine. To process a standing query on a dynamic graph, the framework spawns a predetermined number of threads. Each thread repeatedly performs two tasks: \textit{(i)} dequeuing and applying topology updates and executing event handlers, and \textit{(ii)} only when the topology event buffer is empty (thus prioritizing topology events), consuming messages from other vertices and executing message handlers. These threads communicate with each other via passing messages through FIFO queues.

\subsubsection{Result Collection} To simulate on-demand queries, the user triggers result collection on observing an event with a timestamp $T > T' + \lambda$, where $\lambda$ is the \textit{query interval} and $T'$ is the timestamp triggering the last collection. For result collection, the framework blocks topology events until the algorithm converges and the result is copied out. We refer to the delay between requesting collection to result collection as \textit{result latency}. Importantly, the system serves on-demand requests, which are not restricted to pre-defined intervals, and it does not know in advance when result collection will be triggered.

\subsubsection{Machine}
For experiments, we use a commodity desktop with a 16-core AMD Ryzen 9 5950x, 128GB of RAM, and an NVMe SSD. We parallelize across all cores, except for the scalability experiment (\RefSection{sec:evaluation:scalability}).

\subsubsection{Dataset}
Unfortunately, few large real-world timestamped graphs are available. We source several large real-world timestamped graphs from Mislove et al. \cite{mislove2008growth, mislove2009online} and our prior work \cite{sallinen2023realtime}. \autoref{tab:graphs} summarizes the graphs evaluated. All are multigraphs, and a weight of $1$ is assigned to edges in unweighted graphs. The graphs are stored as plain-text event logs sorted by timestamps. We refer to one day in the event log as one \textit{dataset day}. We restrict the Ethereum graph to transfers of at least 0.1 ETH to create a sub-graph that fits in memory on our single desktop machine.

For the maximum flow problem, the choice of source vertex and sink vertex $(s,t)$ has a significant impact on the results. Therefore, in our experiments, we evaluate \textit{extreme/hardest} cases: we choose sinks as the most popular vertices pre-determined with PageRank (an algorithm representing a likelihood to ``arrive'' at a given vertex) \cite{sallinen2023realtime}, and sources similarly as the most popular vertices on the transpose graph. 

\begin{table}
\setlength{\tabcolsep}{3pt}
\centering\resizebox{\linewidth}{!}{\begin{tabular}{l|rrrrrrrrrr}
\toprule
Graph & $|V|$  & $|E|$ & $\max(in)$ & $\max(out)$ & $\Delta$TS(days) & $\Delta E$/day & Wt. \\
\midrule
Eth-transfers (0.1 min) \cite{sallinen2023realtime} & 82.1M & 475M  & 21M & 19M & 2,772 & 171K &\checkmark \\
Hive-comments \cite{sallinen2023realtime} & 0.7M   & 80M   & 700K & 2.7M & 2,362 & 34K & x \\
Wikipedia-growth \cite{mislove2009online} & 1.9M   & 40M   & 200K & 7K & 2,246 & 18K & x \\
Flickr-growth \cite{mislove2008growth} & 2.3M   & 33M   & 21K & 26K & 179 & 184K & x \\
\bottomrule
\end{tabular}}
\vspace{4pt}
\caption{\label{tab:graphs}Properties of the graphs used in evaluation: total vertex and edge counts, max vertex in- and out-degree, range of timestamps, average number of events per dataset day, and whether edges have weights.}
\end{table}

\subsection{\label{sec:evaluation:staturation}Saturation Experiment}
\vspace{-2pt}
\noindent
This experiment aims to estimate the maximum throughput our solution is able to achieve under various query intervals. For this experiment, we let the framework ingest events as fast as it can and report the ingestion rates (i.e. throughput). Note that the framework prioritizes topological events over algorithmic messages and blocks topological events when there is a pending query.

\begin{figure}
    \centering
    \includegraphics[width=\linewidth]{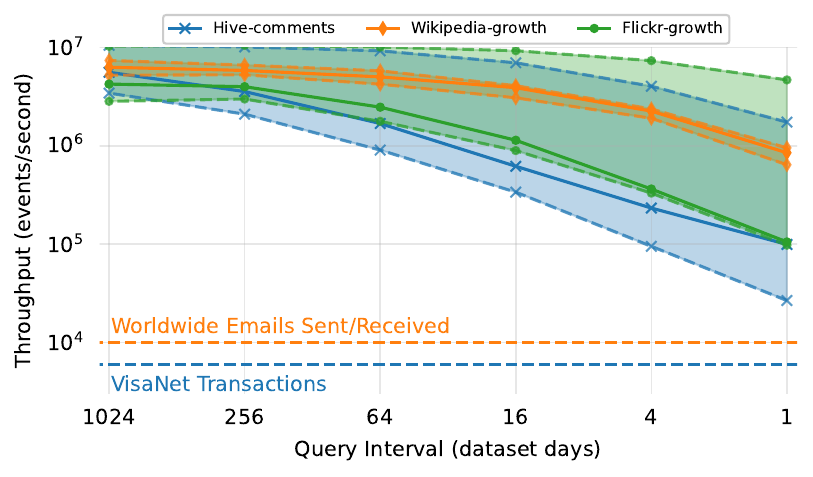}
    \caption{Throughput Under Different Query Intervals. For each graph, the solid line plots the median throughput achieved during graph evolution (y-axis, log scale), for different query intervals (x-axis, log scale). Experiments are repeated for the top 10 $(s,t)$; the dashed lines and the shaded area show the range of throughputs. For comparison, the figure also presents estimated event rates of real-world systems: VisaNet transactions (6,000 e/s) \cite{visa2022}, and worldwide emails sent/received (10,000 e/s) \cite{radicati2022email}.}
    \label{fig:saturation}
    \vspace{-4pt}
\end{figure}

\autoref{fig:saturation} plots the throughput for different graphs under different query intervals. To highlight the magnitude of the achieved throughput, the plot also presents the estimated event rates for a few common real-world systems: the VISA transaction processing system and email. We note a few observations.

First, our solution is able to process events at a rate beyond the rates of common real-world systems, even for queries with a fine granularity of change (e.g., 1 dataset day, implying a result collection request is inserted after each couple of thousand topology events) and even on a commodity desktop. We emphasize that we evaluate \textit{extreme} cases, and our experience shows that our algorithm is able to process with much higher performance if $s$ and $t$ are chosen randomly rather than from the set of most popular vertices.

\begin{figure}
    \centering
    \includegraphics[width=\linewidth]{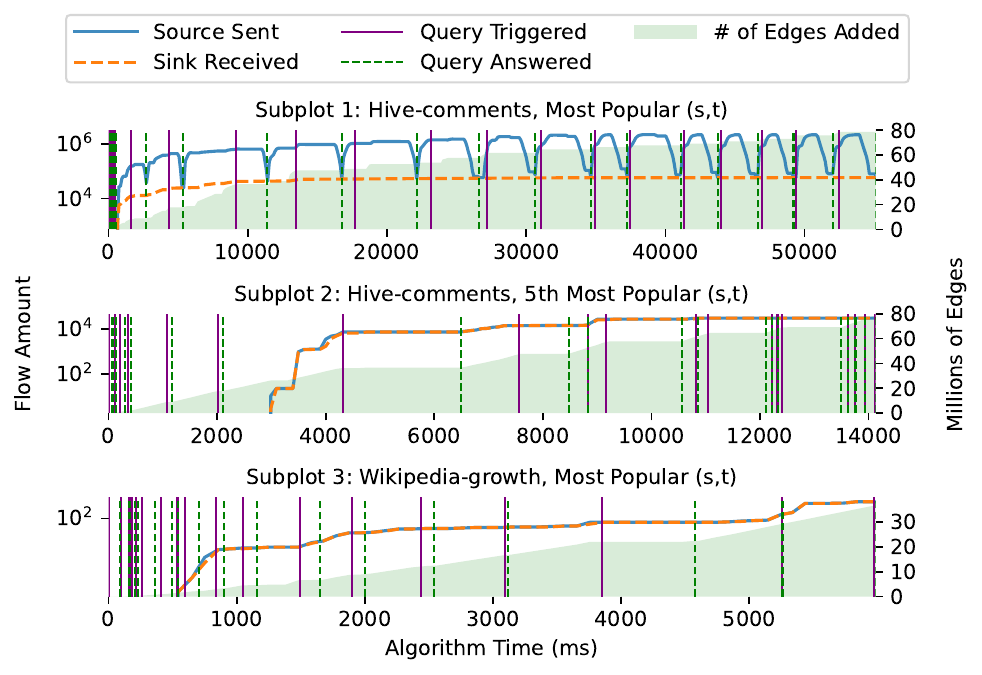}
    \caption{Visual Comparison of Algorithm Progress. For different graphs and $(s,t)$, plotted is the amount of flow (left y-axis, log scale) and the growth of the graph in edges (right y-axis), over time as the graph evolves (x-axis). 128-day query interval.}
    \label{fig:progress}
\end{figure}

Second, throughput depends on the query frequency. As expected, for smaller intervals (i.e. higher query frequencies), our solution tends to have lower throughput as additional resources are allocated to solve the query and collect the result. Upon inspection, we find that handling edge additions sometimes involves an expensive process: the affected vertices may pull more flow from the source than the actual increase in the maximum flow, and the extra flow pulled from the source must be returned for convergence. With larger intervals, the impact of this scenario is amortized as more edge updates are processed before a full convergence. Further, we note this scenario only happens in some datasets: as visualized in \autoref{fig:progress}, while the top $(s,t)$ in the Hive-comments graph is subject to this scenario (Subplot 1), the Wikipedia-growth graph is not (Subplot 3).

Finally, on the same graph, throughput may also have a large variation for different $(s,t)$, up to orders of magnitude. This is because the maximum flow between different $(s,t)$ may involve different sections of the graph, and some sections are not subject to the undesirable scenario mentioned earlier (demonstrated by Subplot 1 and Subplot 2 in \autoref{fig:progress}).

\subsection{\label{sec:evaluation:scalability}Scalability Analysis}
\begin{figure}
    \centering
    \includegraphics[width=0.9\linewidth]{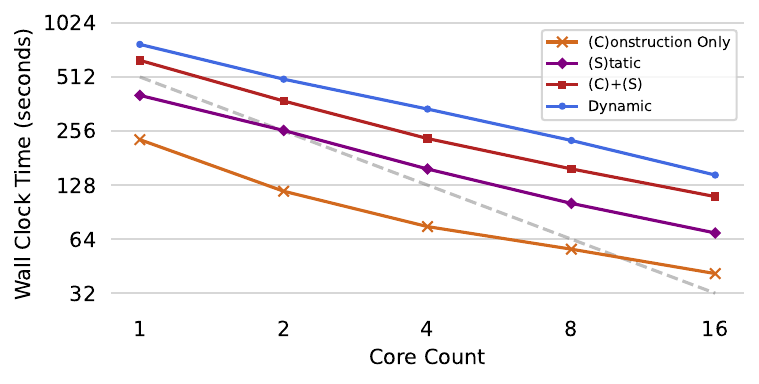}
    \caption{Strong Scaling. The figure plots the total runtime (y-axis, log scale) while varying the number of CPU cores (x-axis, log scale). \textit{Dynamic}: Ingesting events with the algorithm running (on-demand collection available at any time). \textit{Construction Only}: Ingesting events only with no algorithm running. \textit{Static}: Solely running the algorithm on the final ingested graph. The experiment uses the most popular $(s,t)$, and one query to compare to the static approach.}
    \label{fig:scalability}
\end{figure}

\vspace{-2pt}
\noindent
\autoref{fig:scalability} shows the results of strong scaling experiments on the Eth-transfers graph. The key takeaways are \textit{(i)} the cost of dynamic solution tracking (blue line) scales similarly to running the algorithm on the static graph (dark red), and \textit{(ii)} there is no visible deterioration in scaling properties for larger core counts for neither topology construction nor algorithm-level processing (for either of dynamic and static approaches). This suggests that our solution has the potential to be deployed on larger machines to improve performance.

\subsection{\label{sec:evaluation:deletion}Performance of Deletions}
\vspace{-2pt}
\noindent
To show that our algorithm handles deletions efficiently, we evaluate total runtime and result latency on graph evolution traces with and without deletions. To generate an event stream with deletions, we emulate a sliding window view in the topology activity (which leads to delete-heavy workloads): for a window size $W$, on observation of an event with timestamp $T$, the event stream generates the deletions of events occurring before $T-W$.

\autoref{tab:deletions} presents the results. The algorithm is able to achieve similar runtime and latency on graphs with deletions injected, when compared to the results on the original incremental-only graphs. For Hive-comments and Flickr-growth, the algorithm is able to achieve lower runtime and latency with deletions. Note two primary drivers affect the results: \textit{(i)} the sliding window approach creates a trace that has nearly 2$\times$ number of events, and \textit{(ii)} the size of the graph that is active (i.e., the subgraph that should be considered for maximum flow) is smaller due to old edges being removed.

\begin{table}
\centering\resizebox{\linewidth}{!}{\begin{tabular}{l|rr|rr|rr}
\multicolumn{7}{c}{\textit{Total Runtime (s)}} \\
\toprule
Graph & 128-a & 128-d & 64-a & 64-d & 32-a & 32-d \\
\midrule
Eth-transfers & 339.70 & 636.54 & 480.18 & 836.96 & 933.19 & 1370.78 \\
Hive-comments     & 32.26 & 22.77 & 48.44 & 29.43 &  77.86 & 40.86 \\
Wikipedia-growth  &  7.20 & 15.48 &  7.47 & 21.53 &   8.35 & 35.69 \\
Flickr-growth     & 11.86 & 10.73 & 13.17 & 12.01 &  18.27 & 15.22 \\
\bottomrule
\multicolumn{7}{c}{} \\
\multicolumn{7}{c}{\textit{Average Result Latency (s)}} \\
\toprule
Graph & 128-a & 128-d & 64-a & 64-d & 32-a & 32-d \\
\midrule
Eth-transfers & 15.18 & 28.04 & 12.24 & 21.44 & 13.38 & 18.73 \\
Hive-comments     &  1.65 & 0.73 & 1.48 & 0.62 & 1.31 & 0.52 \\
Wikipedia-growth  &  0.33 & 1.05 & 0.22 & 1.09 & 0.12 & 1.18 \\
Flickr-growth     &  3.49 & 2.82 & 3.26 & 2.94 & 3.25 & 2.47 \\
\bottomrule
\end{tabular}}
\vspace{4pt}
\caption{\label{tab:deletions}Total Runtime and Result Latency on Graphs with and without Deletions. Columns: Query interval (in dataset days) and whether the event stream is (a)dd-only or with (d)eletions (generated using a sliding window of size $W=120\text{ days}$). Average over the top 10 $(s,t)$ pairs.}
\end{table}

\subsection{\label{sec:evaluation:rate-limiting}Result Latency vs. Ingestion Rate}
\noindent
When evaluating streaming frameworks, it is common to simulate a stream of incoming events by reading from a file, and letting the framework ingest events as fast as possible. However, when deployed in practice, the incoming event rate is bound by the rate of change in the physical system tracked, and the dynamic graph processing system will usually be provisioned to have more resources than is strictly needed. Thus, we investigate this more likely scenario: the offered event rate is below the framework's maximum throughput.

\begin{figure}
    \centering
    \includegraphics[width=0.9\linewidth]{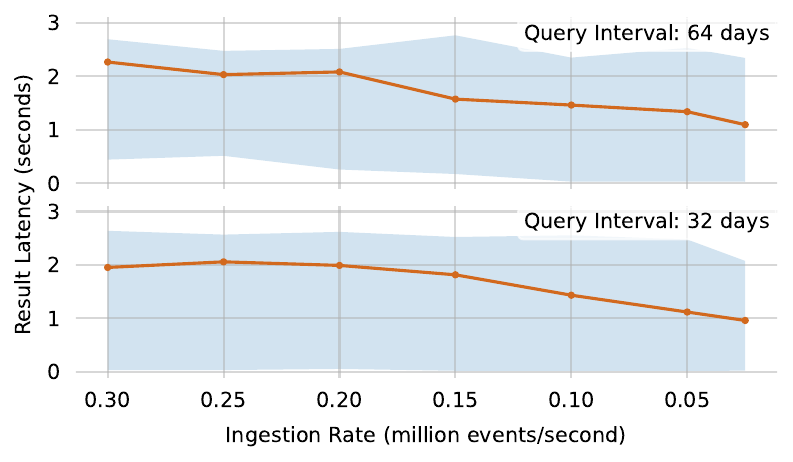}
    \caption{Latency Under Different Event Rates and Query Intervals. The lines plot the median result latency (y-axis) for a given restricted ingestion rate (x-axis). The shaded areas show the range from 20 to 80 percentiles. Graph: Hive-comments (most popular $(s,t)$). }
    \label{fig:rate-limiting}
\end{figure}

To this end, we control the offered event rates and measure the query latency. The result is presented in \autoref{fig:rate-limiting}, which shows the query latency on the Hive-comments graph under different ingestion rates. As our system is not subject to batch constraints (i.e., it does not wait for the entire batch to be ingested before making progress on the algorithm), it is able to leverage the ``free'' CPU time available at lower offered event rates to produce results with lower latency. This demonstrates its ability to process real-world events in real-time.

\subsection{\label{sec:evaluation:stability}Solution Stability}
\noindent
An evaluation criterion for algorithms on evolving structures is \textit{solution stability}. While a static approach recomputes a solution from scratch each time one is demanded, our dynamic approach maintains the algorithm state as the graph evolves, and thus is able to produce 'similar' solutions. In other words, although there may exist multiple sets of paths that result in the same maximum flow, the dynamic approach attempts to only \textit{modify} the prior result rather than to simply produce a new valid, but arbitrary, result. As max-flow is often used for resource allocation or provisioning, a stable algorithm that leads to less resource migration during evolution is desirable.

To analyze the solution stability of the approaches on dynamic graphs, we use the Wikipedia-growth graph as a case study. For each query result, we extract the set of vertices involved in the maximum flow, and report the percentage of these vertices that also appeared in the result of the prior query.

\begin{figure}
    \centering
    \includegraphics[width=\linewidth]{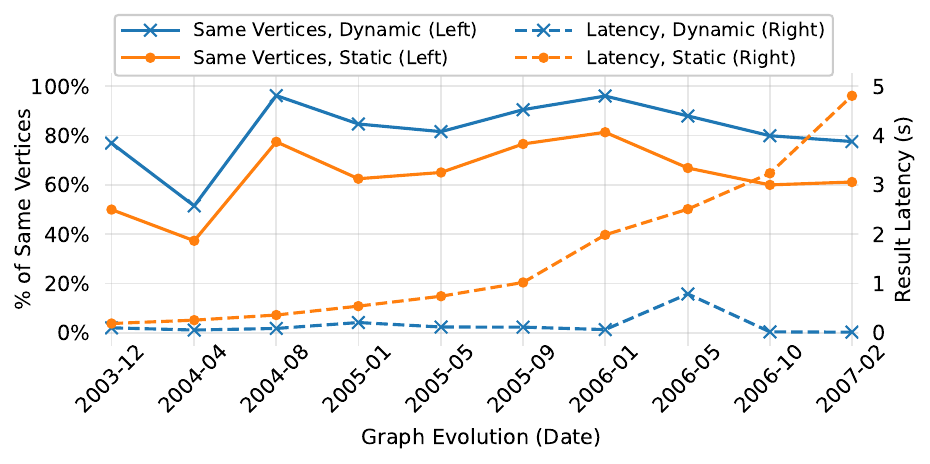}
    \caption{Stability and Result Latency Over Time. The graph evolution (x-axis) is plotted against the percentage of vertices in the current result that also appeared in the previous result (y-axis, left) and result latency (y-axis, right). Graph: Wikipedia-growth.}
    \label{fig:stability}
\end{figure}

\autoref{fig:stability} plots the stability and result latency for each query as the graph evolves, using a static approach (our algorithm running on the same topology, but from scratch on each snapshot) as a baseline. Compared to the static approach, the dynamic approach provides advantages on two fronts: \textit{(i)} it is able to produce significantly more stable results over time, and \textit{(ii)} it is able to produce results with lower latency.
In summary, the dynamic solution is both \textit{timely} and \textit{effective}.

\section{\label{sec:related-work}Related Work}

\subsection{\label{sec:related-work:max-flow}Maximum Flow}
\noindent 
The maximum flow problem has received much attention, with numerous algorithms focused on solving it on \textit{static} graphs (e.g., \cite{harris1955fundamentals, ford1956maximal, goldberg1988new, edmonds1972theoretical, dinitz2006dinitz, cheriyan1988analysis, chen2022maximum, cruz2023survey}). For \textit{dynamic} graphs, there exist only a handful of solutions: Hanauer et al. \cite{hanauer2022recent} survey the sequential ones (e.g., \cite{greco2017incremental, goldberg2015faster, kumar2003incremental, kohli2007dynamic}), and we summarize parallel ones below. 

Although there exist several parallel and dynamic push-relabel algorithms, all have various shortcomings: some impose restrictions on the graph topology (e.g., to directed acyclic graphs), most make assumptions about graph evolution (e.g., only one supports vertex additions), and none of them is designed for highly dynamic graphs.

Zhou proposed an algorithm based on the push-relabel algorithm \cite{goldberg1988new} and adapts to topological updates by resetting the algorithmic state of all vertices \cite{zhou1996self}. Ghosh et al.'s algorithm lets the sink pull from the source and is designed only for directed acyclic graphs \cite{ghosh1997self}. Hong and Prasanna proposed an algorithm based on push-relabel for task allocation \cite{hong2004distributed}. Their algorithm handles capacity changes by increasing the height of the source by $3|V|$ for every change in the capacity of an edge. Pham et al. \cite{pham2006adaptive} and Homayounnejad et al. \cite{homa2011asynchronous, homa2015efficient} proposed adaptive algorithms based on push-relabel, but we found them produce inaccurate results \cite{luo2023going}. Importantly, the algorithms above are designed with a mindset of computing max-flow in a small communication network (up to thousands of nodes) with the objective of decentralization, rather than performance. Recently, Khatri et al. proposed a \textit{pull-relabel} algorithm \cite{khatri2022scaling}, which is a symmetric counterpart of the push-relabel algorithm. Their algorithm runs push-relabel and pull-relabel in turn repeatedly to handle edge modifications.

Of these parallel algorithms, only the solution proposed by Ghosh et al. \cite{ghosh1997self} supports the addition of vertices. Further, all require shared state between vertices; and, most importantly, none are designed for highly dynamic graphs -- they are designed for, and evaluated with, thousands of \textit{total} changes, as opposed to thousands to millions of changes \textit{per second} as we explored. A new algorithm was hence required to process large dynamic graphs with frequent changes in both vertices and edges.

Our prior solution \cite{luo2023going} aims to handle highly dynamic graphs, but it faces dramatic performance degradation when the proportion of deletions increases and lacks critical performance optimizations. The algorithm we propose in this paper is free of this limitation (as shown in \RefSection{sec:evaluation:deletion}), and has a significantly better performance.

\subsection{Graph Streaming Frameworks}
\noindent 
A recent taxonomy of dynamic graph processing frameworks by Besta et al. \cite{besta2020practice} observes that most of the existing systems operate with \textit{batches}: the system ingests batches of graph updates, and provides the algorithm with a static snapshot of an evolving graph after each batch; the algorithm then computes the solution for this snapshot, often warm-starting with results from prior snapshots. The process of ingesting events and executing the algorithm can be either \textit{interleaved} (event ingestion stops when the algorithm is running) \cite{graphbolt} or \textit{pipelined} (event ingestion continues when the algorithm is operating on a static snapshot) \cite{graphone}.

Instead, we develop our algorithm on top of a flexible, scalable computational model that supports retrieving results on demand (i.e., it is not subject to batch constraints) (details in \RefSection{sec:model} and \cite{sallinen2023realtime}). The system processes graph updates and algorithm messages concurrently, and the algorithm reacts to topology changes right after the graph is updated (as opposed to waiting until the current batch is fully ingested). The model is also shared-nothing (vertices communicate only via explicit message passing) to preserve scalability for distributed settings. This model offers several advantages: \textit{(i)} it is scalable (e.g., has been deployed to process a 257-billion edge web graph on 1,000s of cores \cite{sallinen2016graph}); \textit{(ii)} it enables the solution of a graph analytics query to be extracted on-demand with fine granularity and low latency as graphs evolve \cite{sallinen2019incremental, sallinen2023realtime}; and \textit{(iii)} it enables to system to utilize slack resources when the incoming event rate is low (as demonstrated in \RefSection{sec:evaluation:rate-limiting}). The main challenge of developing on this model is designing algorithms that support asynchronicity and concurrency -- there is no shared state or direct synchronization between vertices and the graph may change when the algorithm is running.

\section{Conclusion}
\noindent
We present a dynamic algorithm for the maximum flow problem, and provide a thorough experimental evaluation of the algorithm with difficult cases on large real-world dynamic graphs. Our algorithm works well on a shared-nothing, asynchronous computational model with concurrent graph updates. The evaluation shows our algorithm obtains a high throughput, supports both additions and deletions of vertices and edges efficiently, matches well to parallel and distributed platforms, and provides results with low latency and high stability.



\bibliographystyle{IEEEtran}
\bibliography{IEEEabrv,src/references}

\begin{thebibliography}{10}
\providecommand{\url}[1]{#1}
\csname url@samestyle\endcsname
\providecommand{\newblock}{\relax}
\providecommand{\bibinfo}[2]{#2}
\providecommand{\BIBentrySTDinterwordspacing}{\spaceskip=0pt\relax}
\providecommand{\BIBentryALTinterwordstretchfactor}{4}
\providecommand{\BIBentryALTinterwordspacing}{\spaceskip=\fontdimen2\font plus
\BIBentryALTinterwordstretchfactor\fontdimen3\font minus
  \fontdimen4\font\relax}
\providecommand{\BIBforeignlanguage}[2]{{%
\expandafter\ifx\csname l@#1\endcsname\relax
\typeout{** WARNING: IEEEtran.bst: No hyphenation pattern has been}%
\typeout{** loaded for the language `#1'. Using the pattern for}%
\typeout{** the default language instead.}%
\else
\language=\csname l@#1\endcsname
\fi
#2}}
\providecommand{\BIBdecl}{\relax}
\BIBdecl

\bibitem{besta2020practice}
M.~Besta, M.~Fischer, V.~Kalavri, M.~Kapralov, and T.~Hoefler, ``Practice of
  streaming processing of dynamic graphs: Concepts, models, and systems,''
  \emph{IEEE Transactions on Parallel and Distributed Systems}, 2021.

\bibitem{aggarwal2014evolutionary}
C.~Aggarwal and K.~Subbian, ``Evolutionary network analysis: A survey,''
  \emph{ACM Computing Surveys (CSUR)}, vol.~47, no.~1, pp. 1--36, 2014.

\bibitem{mcgregor2014graph}
A.~McGregor, ``Graph stream algorithms: a survey,'' \emph{ACM SIGMOD Record},
  vol.~43, no.~1, pp. 9--20, 2014.

\bibitem{graphone}
P.~Kumar and H.~H. Huang, ``Graphone: A data store for real-time analytics on
  evolving graphs,'' \emph{ACM Transactions on Storage (TOS)}, vol.~15, no.~4,
  pp. 1--40, 2020.

\bibitem{graphbolt}
M.~Mariappan and K.~Vora, ``Graphbolt: Dependency-driven synchronous processing
  of streaming graphs,'' in \emph{Proceedings of the Fourteenth EuroSys
  Conference 2019}, 2019, pp. 1--16.

\bibitem{sallinen2016graph}
S.~Sallinen, K.~Iwabuchi, S.~Poudel, M.~Gokhale, M.~Ripeanu, and R.~Pearce,
  ``Graph colouring as a challenge problem for dynamic graph processing on
  distributed systems,'' in \emph{SC'16: Proceedings of the International
  Conference for High Performance Computing, Networking, Storage and
  Analysis}.\hskip 1em plus 0.5em minus 0.4em\relax IEEE, 2016, pp. 347--358.

\bibitem{sallinen2019incremental}
S.~Sallinen, R.~Pearce, and M.~Ripeanu, ``Incremental graph processing for
  on-line analytics,'' in \emph{2019 IEEE International Parallel and
  Distributed Processing Symposium (IPDPS)}.\hskip 1em plus 0.5em minus
  0.4em\relax IEEE, 2019, pp. 1007--1018.

\bibitem{west2001introduction}
D.~B. West \emph{et~al.}, \emph{Introduction to graph theory}.\hskip 1em plus
  0.5em minus 0.4em\relax Prentice hall Upper Saddle River, 2001, vol.~2.

\bibitem{Ahuja1993network}
R.~K. Ahuja, T.~L. Magnanti, and J.~B. Orlin, \emph{Network Flows: Theory,
  Algorithms, and Applications}.\hskip 1em plus 0.5em minus 0.4em\relax USA:
  Prentice-Hall, Inc., 1993.

\bibitem{harris1955fundamentals}
T.~Harris and F.~Ross, ``Fundamentals of a method for evaluating rail net
  capacities,'' RAND CORP SANTA MONICA CA, Tech. Rep., 1955.

\bibitem{homa2015efficient}
S.~Homayounnejad and A.~Bagheri, ``An efficient distributed max-flow algorithm
  for wireless sensor networks,'' \emph{Journal of Network and Computer
  Applications}, vol.~54, pp. 20--32, 2015.

\bibitem{flake2000efficient}
G.~W. Flake, S.~Lawrence, and C.~L. Giles, ``Efficient identification of web
  communities,'' in \emph{Proceedings of the sixth ACM SIGKDD international
  conference on Knowledge discovery and data mining}, 2000, pp. 150--160.

\bibitem{imafuji2004finding}
N.~Imafuji and M.~Kitsuregawa, ``Finding web communities by maximum flow
  algorithm using well-assigned edge capacities,'' \emph{IEICE transactions on
  Information and Systems}, vol.~87, no.~2, pp. 407--415, 2004.

\bibitem{saito2007large}
H.~Saito, M.~Toyoda, M.~Kitsuregawa, and K.~Aihara, ``A large-scale study of
  link spam detection by graph algorithms,'' in \emph{Proceedings of the 3rd
  international workshop on Adversarial information retrieval on the web},
  2007, pp. 45--48.

\bibitem{jensen2022review}
P.~M. Jensen, N.~Jeppesen, A.~B. Dahl, and V.~A. Dahl, ``Review of serial and
  parallel min-cut/max-flow algorithms for computer vision,'' \emph{IEEE
  Transactions on Pattern Analysis and Machine Intelligence}, vol.~45, no.~2,
  pp. 2310--2329, 2022.

\bibitem{boykov2001experimental}
Y.~Boykov and V.~Kolmogorov, ``An experimental comparison of min-cut/max-flow
  algorithms for energy minimization in vision,'' in \emph{Energy Minimization
  Methods in Computer Vision and Pattern Recognition}, M.~Figueiredo,
  J.~Zerubia, and A.~K. Jain, Eds.\hskip 1em plus 0.5em minus 0.4em\relax
  Berlin, Heidelberg: Springer Berlin Heidelberg, 2001, pp. 359--374.

\bibitem{tran2009sybil}
D.~N. Tran, B.~Min, J.~Li, and L.~Subramanian, ``Sybil-resilient online content
  voting.'' in \emph{NSDI}, vol.~9, no.~1, 2009, pp. 15--28.

\bibitem{krishnamurthy1984improved}
Krishnamurthy, ``An improved min-cut algonthm for partitioning vlsi networks,''
  \emph{IEEE Transactions on computers}, vol. 100, no.~5, pp. 438--446, 1984.

\bibitem{hong2004distributed}
B.~Hong and V.~K. Prasanna, ``Distributed adaptive task allocation in
  heterogeneous computing environments to maximize throughput,'' in \emph{18th
  International Parallel and Distributed Processing Symposium, 2004.
  Proceedings.}\hskip 1em plus 0.5em minus 0.4em\relax IEEE, 2004, pp. 52--.

\bibitem{zhou1996self}
Y.~Zhou, ``A self-stabilizing distributed maximum flow algorithm,'' Ph.D.
  dissertation, University of Nevada, Las Vegas, 1996.

\bibitem{ghosh1997self}
S.~Ghosh, A.~Gupta, and S.~V. Pemmaraju, ``A self-stabilizing algorithm for the
  maximum flow problem,'' \emph{Distributed Computing}, vol.~10, no.~4, pp.
  167--180, 1997.

\bibitem{pham2006adaptive}
T.~L. Pham, M.~Bui, I.~Lavallee, and S.~H. Do, ``An adaptive distributed
  algorithm for the maximum flow problem in the underlying asynchronous
  network,'' in \emph{2006 International Conference onResearch, Innovation and
  Vision for the Future}, 2006, pp. 187--194.

\bibitem{homa2011asynchronous}
S.~Homayounnejad, A.~Bagheri, and A.~Ghebleh, ``Aaa: Asynchronous adaptive
  algorithm to solve max-flow problem in wireless sensor networks,'' in
  \emph{2011 Proceedings of the 34th International Convention MIPRO}.\hskip 1em
  plus 0.5em minus 0.4em\relax IEEE, 2011, pp. 440--445.

\bibitem{khatri2022scaling}
J.~Khatri, A.~Samar, B.~Behera, and R.~Nasre, ``Scaling the maximum flow
  computation on gpus,'' \emph{International Journal of Parallel Programming},
  vol.~50, no. 5-6, pp. 515--561, 2022.

\bibitem{luo2023going}
J.~Luo, S.~Sallinen, and M.~Ripeanu, ``Going with the flow: Real-time max-flow
  on asynchronous dynamic graphs,'' in \emph{Proceedings of the 6th Joint
  Workshop on Graph Data Management Experiences \& Systems (GRADES) and Network
  Data Analytics (NDA)}, 2023, pp. 1--11.

\bibitem{sallinen2023realtime}
S.~Sallinen, J.~Luo, and M.~Ripeanu, ``Real-time pagerank on dynamic graphs,''
  in \emph{2023 ACM International Symposium on High-Performance Parallel and
  Distributed Computing (HPDC)}, 2023.

\bibitem{ford1956maximal}
L.~R. Ford and D.~R. Fulkerson, ``Maximal flow through a network,''
  \emph{Canadian Journal of Mathematics}, vol.~8, p. 399–404, 1956.

\bibitem{goldberg1988new}
A.~V. Goldberg and R.~E. Tarjan, ``A new approach to the maximum-flow
  problem,'' \emph{J. ACM}, vol.~35, no.~4, p. 921–940, oct 1988.

\bibitem{cherkassky1997implementing}
B.~V. Cherkassky and A.~V. Goldberg, ``On implementing the push—relabel
  method for the maximum flow problem,'' \emph{Algorithmica}, vol.~19, pp.
  390--410, 1997.

\bibitem{reza2020highly}
\BIBentryALTinterwordspacing
T.~A. Reza, G.~D. Sanders, K.~Iwabuchi, R.~A. Pearce, and U.~N. N.~S.
  Administration, ``Highly asynchronous visitor queue graph toolkit, version
  0.2,'' 9 2020. [Online]. Available: \url{https://www.osti.gov/biblio/1737365}
\BIBentrySTDinterwordspacing

\bibitem{pearce2013scaling}
R.~Pearce, M.~Gokhale, and N.~M. Amato, ``Scaling techniques for massive
  scale-free graphs in distributed (external) memory,'' in \emph{Parallel \&
  Distributed Processing (IPDPS), 2013 IEEE 27th International Symposium
  on}.\hskip 1em plus 0.5em minus 0.4em\relax IEEE, 2013, pp. 825--836.

\bibitem{goldberg2015faster}
A.~V. Goldberg, S.~Hed, H.~Kaplan, P.~Kohli, R.~E. Tarjan, and R.~F. Werneck,
  ``Faster and more dynamic maximum flow by incremental breadth-first search,''
  in \emph{Algorithms-ESA 2015: 23rd Annual European Symposium, Patras, Greece,
  September 14-16, 2015, Proceedings}.\hskip 1em plus 0.5em minus 0.4em\relax
  Springer, 2015, pp. 619--630.

\bibitem{kulkarni2009much}
M.~Kulkarni, M.~Burtscher, R.~Inkulu, K.~Pingali, and C.~Cas{\c{c}}aval, ``How
  much parallelism is there in irregular applications?'' \emph{ACM sigplan
  notices}, vol.~44, no.~4, pp. 3--14, 2009.

\bibitem{mislove2008growth}
A.~Mislove, H.~S. Koppula, K.~P. Gummadi, P.~Druschel, and B.~Bhattacharjee,
  ``Growth of the flickr social network,'' in \emph{Proceedings of the first
  workshop on Online social networks}, 2008, pp. 25--30.

\bibitem{mislove2009online}
A.~E. Mislove, \emph{Online social networks: measurement, analysis, and
  applications to distributed information systems}.\hskip 1em plus 0.5em minus
  0.4em\relax Rice University, 2009.

\bibitem{visa2022}
\BIBentryALTinterwordspacing
{VISA INC.}, ``Visa annual report 2022,'' 2022. [Online]. Available:
  \url{https://s29.q4cdn.com/385744025/files/doc_downloads/2022/Visa-Inc-Fiscal-2022-Annual-Report.pdf}
\BIBentrySTDinterwordspacing

\bibitem{radicati2022email}
\BIBentryALTinterwordspacing
{The Radicati Group, Inc.}, ``Email statistics report, 2022-2026,'' November
  2022. [Online]. Available: \url{https://www.radicati.com/?p=17936}
\BIBentrySTDinterwordspacing

\bibitem{edmonds1972theoretical}
J.~Edmonds and R.~M. Karp, ``Theoretical improvements in algorithmic efficiency
  for network flow problems,'' \emph{Journal of the ACM (JACM)}, vol.~19,
  no.~2, pp. 248--264, 1972.

\bibitem{dinitz2006dinitz}
Y.~Dinitz, ``Dinitz’algorithm: The original version and even’s version,''
  in \emph{Theoretical Computer Science: Essays in Memory of Shimon
  Even}.\hskip 1em plus 0.5em minus 0.4em\relax Springer, 2006, pp. 218--240.

\bibitem{cheriyan1988analysis}
J.~Cheriyan and S.~N. Maheshwari, ``Analysis of preflow push algorithms for
  maximum network flow,'' in \emph{Foundations of Software Technology and
  Theoretical Computer Science}, K.~V. Nori and S.~Kumar, Eds.\hskip 1em plus
  0.5em minus 0.4em\relax Berlin, Heidelberg: Springer Berlin Heidelberg, 1988,
  pp. 30--48.

\bibitem{chen2022maximum}
L.~Chen, R.~Kyng, Y.~P. Liu, R.~Peng, M.~P. Gutenberg, and S.~Sachdeva,
  ``Maximum flow and minimum-cost flow in almost-linear time,'' in \emph{2022
  IEEE 63rd Annual Symposium on Foundations of Computer Science (FOCS)}.\hskip
  1em plus 0.5em minus 0.4em\relax IEEE, 2022, pp. 612--623.

\bibitem{cruz2023survey}
O.~Cruz-Mej{\'\i}a and A.~N. Letchford, ``A survey on exact algorithms for the
  maximum flow and minimum-cost flow problems,'' \emph{Networks}, 2023.

\bibitem{hanauer2022recent}
K.~Hanauer, M.~Henzinger, and C.~Schulz, ``Recent advances in fully dynamic
  graph algorithms--a quick reference guide,'' \emph{ACM Journal of
  Experimental Algorithmics}, vol.~27, pp. 1--45, 2022.

\bibitem{greco2017incremental}
S.~Greco, C.~Molinaro, C.~Pulice, and X.~Quintana, ``Incremental maximum flow
  computation on evolving networks,'' in \emph{Proceedings of the Symposium on
  Applied Computing}, 2017, pp. 1061--1067.

\bibitem{kumar2003incremental}
S.~Kumar and P.~Gupta, ``An incremental algorithm for the maximum flow
  problem,'' \emph{Journal of Mathematical Modelling and Algorithms}, vol.~2,
  pp. 1--16, 2003.

\bibitem{kohli2007dynamic}
P.~Kohli and P.~H. Torr, ``Dynamic graph cuts for efficient inference in markov
  random fields,'' \emph{IEEE transactions on pattern analysis and machine
  intelligence}, vol.~29, no.~12, pp. 2079--2088, 2007.

\end{thebibliography}

\end{document}